\documentclass[a4paper, twocolumn, 12pt]{article}
\usepackage[utf8]{inputenc}
\usepackage{cite}
\usepackage[centertags]{amsmath}
\usepackage{amsfonts}
\usepackage{textcomp}
\usepackage{amssymb}
\usepackage{amsthm}
\usepackage[a4paper,left=1.5cm,right=1cm]{geometry}
\usepackage{graphicx}
\usepackage{hyperref}
\usepackage{newlfont}
\usepackage{mathtools, cuted}
\usepackage{lipsum, color}
\usepackage{multicol}
\usepackage{xcolor, soul}
\sethlcolor{green}
\newcommand{\dbar}{d\hspace*{-0.08em}\bar{}\hspace*{0.1em}}

\usepackage{xtocinc} 
\usepackage[active]{srcltx} 
\title{\vspace{-2.5cm}\textbf{Finite-time thermodynamic process of a two-level quantum heat engine}}

\author{\small Yigermal Bassie $^a$,Tibebe Birhanu$^b$,  Yoseph Abebe $^c$ and Admasu Abawari$^d$ \\ 
\small $^a$Department of Physics, Wolkite University, Wolkite, Ethiopia.\\
\small $^b$Department of Physics,  University of Gondar, Gondar, Ethiopia.\\
 \small $^c$Department of Physics, Debre Markos University, Debre Markos, Ethiopia.\\
 \small $^d$Department of Physics, Addis Ababa University, Addis Ababa, Ethiopia.
}

\begin{document}
\twocolumn[
  \begin{@twocolumnfalse}
\maketitle
\begin{abstract}
\noindent In this paper, we consider a  model of two-level quantum heat engine to investigate the explicit analytic expression for the thermodynamics quantities in different condition under the finite-time operation. In this engine, the working substance is composed of a spin-half particles immersed in a magnetic field. The finite-time thermodynamic processes consisting of two quantum adiabatic and two quantum isothermal processes. This processes working between two heat reservoirs with an inverse temperatures $\beta_{1}$ and $\beta_{2}$ ($<\beta_{1}$). In this processes, we obtain the work, heat, power and efficiency at maximum power output of the model. Our result of the efficiency at maximum power agree with the universal value in the first order of Carnot efficiency.\\

\noindent keywords \textit{Quantum heat engine, efficiency at maximum power, two-level system, finite-time processes}
\end{abstract}
\end{@twocolumnfalse}
  ]
\noindent A finite-time process in thermodynamics is a process that transforms a given initial state into a given final state in a finite amount of time. This process is optimal if it produces the lowest amount of entropy production.
The condition of a finite time is crucial, because quasi-static processes, which require infinitely slow driving, do not generate any entropy at all. Such processes have been studied in macroscopic systems under the realm of finite-time thermodynamics.  \cite{andresen2011current}. For small systems in contact with a thermal environment and thus following a stochastic dynamics, optimal finite-time processes were shown to have an inevitable thermodynamic cost that scales asymptotically like the inverse of the allocated time \cite{sekimoto1997complementarity}, \cite{schmiedl2007optimal}. This scaling was later shown to be the exact minimal entropy production for any finite time for an underlying Langevin dynamics \cite{schmiedl2007efficiency}, \cite{aurell2012boundary}. For system with discrete state space undergoing a master equation dynamics, this scaling holds asymptotically as several case studies have shown \cite{esposito2010finite}, \cite{diana2013finite}, \cite{zulkowski2014optimal}. 
In the linear response regime, an appealing systematic
theory for the optimal driving involves geometric concepts like the thermodynamic length \cite{crooks2007work}, \cite{sivak2012thermodynamic}, \cite{sivak2016thermodynamic}. For an effective two-state system, a prominent experimental application of optimal protocols is the minimal cost of erasing a bit in a finite-time extension of the Landauer bound
\cite{sivak2016thermodynamic}, \cite{jun2014high}, \cite{proesmans2020finite}.
\\ 
The concept of thermodynamics has been developed from the analysis of heat engines performance. Carnot invented an idealized mathematical model of  heat engines called the Carnot cycle and proved that there exists a maximum efficiency of all heat engines, which is given by Carnot efficiency. This efficiency is a central cornerstone of thermodynamics. It states that the efficiency of a reversible Carnot heat engine attain the maximum possible work for a given temperature of the hot $(T_{h})$ and cold ($T_{c}$) reservoirs but generates zero power because it is an infinitely slow operation. The efficiency ($\eta_{c}=1-\frac{T_{c}}{T_{h}} $) of the Carnot cycle is the upper bound on the efficiency at which real heat engines are unrealistically high. The practical implications are more limited, since the upper limit $\eta_{c}$ is only reached for engines that operate reversibly. One of the important questions is what will be the efficiency at maximum power of a system that operating in finite time. In a groundbreaking work, Curzon and Ahlborn \cite{curzon1975efficiency} obtained this efficiency for the Carnot engine by optimizing the Carnot cycle with respect to power rather than efficiency, which is given by Curzon–Ahlborn efficiency, $\eta_{CA}$
\begin{eqnarray}\label{eq1}
 \eta_{CA} =1-\sqrt{\frac{T_{c}}{T_{h}}} = \frac{\eta_{c}}{2}+ \frac{\eta_{c}^{2}}{8} +\vartheta(\eta^{3}_{c})  .
\end{eqnarray}
This efficiency used to seek a more realistic upper bound on the efficiency of a heat engine in the endoreversible approximation \cite{curzon1975efficiency},\cite{callen1985thermodynamics} (taking into account the dissipation only in the heat transfer process). Currently, it has been shown that the Curzon-Ahlborn efficiency is an exact consequence of linear irreversible thermodynamics when operating under conditions of strong coupling between the heat flux and the work \cite{van2005thermodynamic}, \cite{de2007collective}, \cite{gomez2006tight}. 
The value of 1/2 for the linear coefficient in Eq. \ref{eq1} is therefore universal for such systems.\\
\noindent In this paper we consider a two-level quantum system with excited (ground) states  and  with  its corresponding eigen-energy values.  Such a system modelled as a spin-half particle in an external magnetic field. It is in contact with a two thermal reservoirs having different inverse temperature $\beta_{1}$ and $\beta_{2}$. The system absorbs heat from the hot reservoir ($\beta_{1}$), do some mechanical work and releases the remaining heat to the cold reservoir ($\beta_{2}$).
We derive analytical expression for the heat absorbed(and released), work done by (and on) per cycle  and the efficiency of a system in a finite time. Furthermore, we explore the behavior of efficiency at maximum power as a function of the Carnot efficiency.\\
\noindent The rest of this paper is organized as follow. In Sec.\eqref{model2}, the model of the system is introduced. In Sec.\eqref{section3} the first law of thermodynamics of the model under finite-time operation explained and thermodynamics quantities also obtained.  In Sec.\eqref{section4} the cyclic process of the the mode of operation described and the condition for finite-time processes is discussed. Section \eqref{section5} analyses the efficiency at maximum power of the model as a function of Carnot efficiency, $\eta_{c}$. In Sec.\eqref{section6}, we summarize and conclude.
\section{The model}\label{model2}
\noindent We consider a two-level quantum system  can be modelled as a spin-half particle in an external magnetic field. 
 In this system,  a working substance consisting of many non interacting spin-half particles and a cycle of engine have two isothermal branches connected by two adiabatic branches, none of which are assumed reversible.
Each of the spin-half particle is thermally coupled to a heat bath of constant inverse temperature of the hot reservoir ($\beta_{1}$) and cold reservoir($\beta_{2}$).
When the system is coupled to a thermal bath and the Hamiltonian is fixed in time, the bath can change the populations of the energy levels. In a steady state, the system reaches a Gibbs state, where the population of the levels is given by\cite{wang2012efficiency}
\begin{equation}\label{one1}
  P_{n,b} = \frac{e^{-\frac{E_{n}}{T_{b}}}}{\sum_{n=1}^{n}{e^{-\frac{E_{n}}{T_{b}}}}×},
\end{equation}
where $n$ is the number of levels and b stands for cold and hot temperature reservoir. 
In next section, we explore the Quantum thermodynamic processes of the model and   evaluate  its thermodynamic quantities such as work, internal energy, heat and entropy.
\section{Quantum thermodynamic process: first law of thermodynamics }\label{section3}
\noindent Consider our two-level system as a spin-half particle in the presence of  an external time dependent magnetic field. The spin-half particles under this field can have two possible discrete energy states values. Since their  magnetic moment prefers to line up either parallel or anti-parallel with the field. This system is placed in contact with a heat reservoir having a constant temperature $\beta$.  
\noindent The system average (internal) energy can be expressed as
\begin{equation}
 \langle E \rangle = \sum_{r}P_{r} E_{r},
\end{equation}
where $P_{r}$ is the probability of finding the system in a microstate r with energy $E_{r}$.
An infinitesimal change $d\langle E \rangle$, in the mean energy can arise in two ways: either (i) in the infinitesimal change in energy level $dE_{r}$ caused by  change in external field at constant probability in $P_{r}$ (this is  carried out instantaneously by detaching  the system from the reservoir) or (ii) in the infinitesimal change in probability, $dP_{r}$, caused by heat exchange of the system with the reservoir at constant external field. In case (i) work will be done by (on) the system depending on the direction of the path. In case (ii) heat will be absorbed (or released) by the system depending on the direction of the path. In thermodynamics, putting the conservation of energy mathematically as,
\begin{equation}\label{firlaw}
d\langle E \rangle = \sum_{r}[P_{r} dE_{r} + E_{r}dP_{r}] = \dbar W+ \dbar Q,
 \end{equation}
where $\dbar W = \sum_{r} P_{r} dE_{r} = P^{e}d\Delta$ is the mean path dependent work done by (on) the system, and
$\dbar Q= \sum_{r} E_{r} dP_{r} = \Delta dP^{e}$ is the mean path dependent heat absorbed (released) by the system.
Note that:  $\Delta$  and $P^{e}$ is the energy gap between the ground state and the excited state and the probability of getting the spin-half particle in the excited state  of the system , respectively and slash on $d$ in Eq.\eqref{firlaw} signifies the path-dependence nature of the process.\\
We determine the entropy as
\begin{equation}\label{entropy}
 S =-\sum_{r}P_{r}\ln P_{r}.
\end{equation}
Finally, using $\dbar Q= \sum_{r} E_{r} dP_{r}$ and Eq. \eqref{entropy} we arrived
\begin{equation}\label{entropy1}
 dS=\frac{\dbar Q}{T}hh
\end{equation}
This relation expressed in Eq. \eqref{entropy1} holds only for quasi-static processes. For in the
case of finite and irreversible processes, it can be additional contributions to the change
in entropy.

\begin{figure}[h]
 \centering
 \includegraphics[width=7cm,height=7cm]{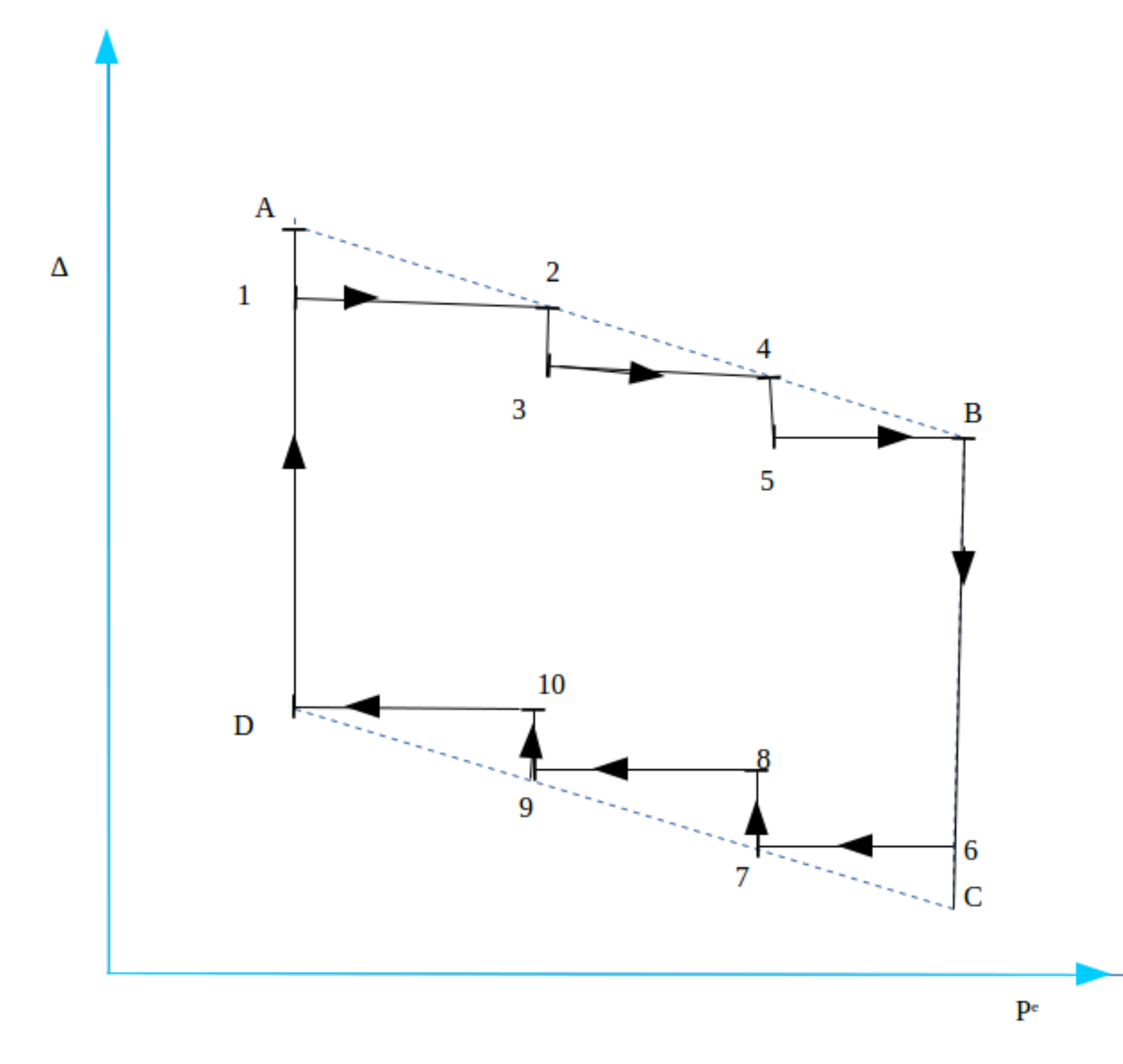}
 \caption{Model heat engine working between two reservoirs at different temperature for finite-time process. Here the horizontal axis $P^{e}$ is the occupation probability in the excited state of the two-level system, and the vertical axis indicates the level spacing of the two-level system. Zig-zag segments represent the finite-time process and broken curves represent quasistatic isothermal processes. $\beta_{1}$ is the inverse temperature of the reservoir at higher temperature; whereas $\beta_{2}$ is the inverse
temperature of the reservoir at lower temperature.}
\label{QHEs}
\end{figure}
\section{Cyclic finite-time process of the system}\label{section4}
\noindent We have consider the finite-time operation of a quantum heat engine whose working substance is composed of
a two-level system. The engine cycle, consisting of two types of processes and working between two heat reservoirs at different temperatures.
We are taking a system of spin-half particles, that are positioned in a fixed space like atoms in a crystal far enough apart to be very weakly interacting with each other. The system is subjected to an external magnetic field. 
Depending on the kind of process to which it is subjected, the system could be attached to a thermal reservoir or detached from it and undergo a change in the field. For example, if the system is exposed to in thermal contact with a heat reservoir of inverse temperature $\beta(=\frac{1}{k_{B}T})$ and let to quasistically change from state A to B, it will trace the isothermal path defined by the relation 
\begin{equation}
 \beta \Delta = \ln\bigg(\frac{1-p^{e}}{p^{e}}\bigg)
\end{equation}
as shown in Fig.\eqref{QHEs}. Here $\Delta(= 2\mu \textbf{B})$ is the energy gap between the ground state and the excited state of a spin-half particle (of magnetic moment $\mu$ ) immersed in an external magnetic field of strength
$B$ while $P^{e}$ is the probability of getting the spin-half particle in the excited state when it is in thermal contact with a heat bath at $\beta$. Let us now discuss each individual branch of finite-time thermodynamic process.
\subsection{Isothermal branch in the finite-time process}
\noindent The two-level quantum heat engine under finite-time operation in the isothermal branch can be described by using the change of energy levels and the number of spin-half particles in the energy state. The two modes of operations in a finite-time process of  the system are clearly shown in the Fig. \eqref{QHEs}.
The first one is an adiabatic change that takes place at constant occupation probability of the excited state. This process involves detaching the system from a reservoir and changing the external magnetic field in a very short time so that only work is done by(on) system; no heat exchange takes place. The second mode of operation in a finite-time process takes place at constant external magnetic field. The detached system from the reservoir during the adiabatic change is attached back to the reservoir during this process. This is a slow process that takes relatively much longer time than the adiabatic change. It proceeds until the system and the reservoir equilibrate. Once the system and the reservoir are in thermal equilibrium, then the first mode of operation takes place followed by the second. These alternate modes of operations occur for a finite number of times which ultimately take the system to a desired final state in a finite time.
 As an illustration shown in Fig.\eqref{QHEs}, we consider two infinitesimal successive processes where the system (i) at first goes from state 2 to state 3 by detaching it from reservoir such that the external field infinitesimally decreased keeping the system probability fixed followed by (ii) a state change from 3 to 4 such that the probability changes infinitesimally while keeping the external field constant. At the end of the second process the system has moved to another state along the isothermal path. In process $2\rightarrow3$, system does work. On the other hand, in the reverse process $3\rightarrow2$ work is done on the system. In process $3\rightarrow4$, system absorbs heat, while in process $4\rightarrow3$ the system released heat. Note that states $2$ and $4$ are lying on the isothermal path. process $1\rightarrow2$ is an instantaneous process, while process $3\rightarrow4$ takes finite time since the system is attached back to $\beta_{1}$ and left there until it equilibrates with reservoir $\beta_{1}$.\\
\noindent In the finite-time process, the spin-half system is connected to the hot reservoir (A to B) with inverse temperature $\beta_{1}$ or the cold reservoir (C to D) with inverse temperature $\beta_{2}$ in the instant time is depicted in Fig.\eqref{QHEs}.
We have considered that the initial state (${P^{e}_{A}, \Delta_{A}}$) reaches in thermal equilibrium within the
hot reservoir at  inverse temperature $\beta_{1}$, when the spin-half system is connected to hot reservoir. When the initial state (${P^{e}_{A}, \Delta_{A}}$) detaches from the hot reservoir, the number of spin-half particles in the exited state remains constant in the first mode of operation, and the change in energy level decreases from $\Delta_{A}$ to $\Delta_{1}$ in a very fast manner. During this abrupt change in energy levels, the system can extract work, with the amount of work done by the system defined as 
\begin{equation}
 W_{A\rightarrow 1} = P^{e}_{A}[\Delta_{1}-\Delta_{A}].
\end{equation}
Using similar ways, the amount of work done by the system in the isothermal processes of A to B can be expressed as
\begin{equation}
\begin{split}
 W_{2\rightarrow 3}& = P^{e}_{2}[\Delta_{3}-\Delta_{2}],
 \text{ } \text{ }  \text { and }\\
 W_{4\rightarrow 5}& = P_{4}^{e}[\Delta_{5}-\Delta_{4}].
\end{split}
 \end{equation}
Therefore, the net work done by a two-level quantum heat engine in the isothermal processes of A to B is given by
\begin{equation}\label{WAB}
 W_{A\rightarrow B} = -\Delta[P^{e}_{A} + P_{2}^{e} +P_{4}^{e}],
\end{equation}
where $\Delta = \frac{\Delta_{A}-\Delta_{B}}{n}$, n is the number of subdivision.\\
\noindent In the second mode of operation, the change of energy level  $\Delta_{1}$  remains constant and the systems is allowed to exchange heat until it relaxes back to thermal equilibrium at state 2, when the system attached to the reservoir. During this process (1 to 2), as seen in the Fig. \eqref{QHEs}, the systems is allowed to absorbed a certain amount of heat
\begin{equation}
 Q_{1\rightarrow2}= \Delta_{1}[P_{2}^{e}-P_{1}^{e}],
\end{equation}
with the hot reservoir $\beta_{1}$. This absorbed heat increases the number of spin-half particles from  the exited state 1($P_{1}^{e},1$)
to the exited state 2 ($P_{2}^{e},2$). Using similar fashion, the amount of absorbed heat during in the 
isothermal processes of A to B can be  obtained as
\begin{equation}
\begin{split}
 Q_{3\rightarrow4}&= \Delta_{4}[P_{4}^{e}-P_{3}^{e}],
\text{ } \text{ }  \text { and }\\
 Q_{5\rightarrow B}&= \Delta_{B}[P_{B}^{e}-P_{5}^{e}].
 \end{split}
\end{equation}
The total amount of heat absorbed by the two-level quantum heat engine in the isothermal processes of A to B is given by 
\begin{equation}
 Q_{A\rightarrow B}= \Delta_{B}P_{4}^{e}-\Delta_{A}P^{e}_{A}+ \Delta [P^{e}_{A}+P_{2}^{e}+P_{4}^{e}].
\end{equation}
During this finite-time process (A to B ), the absorbed heat is positive, where as work done by the system
is negative since the system looses its internal energy.\\
\noindent In the other branch of isothermal process of C to D as depicted in Fig.\eqref{QHEs}, when the systems allows to exchange heat with the cold reservoir at a inverse temperature $\beta_{2}$, after the end of adiabatic process of B to C but thermal equilibrium should be reached within the cold reservoir before the next adiabatic change begins, when the number of spin-half particles in the excited state decreases. By increasing the external magnetic field adiabatically,
it is possible to increase the energy level of the system from $ \Delta_{C} $ to $ \Delta_{6} $ and further continue to the processes $6\rightarrow7$, $8\rightarrow9$, and $10\rightarrow D$ while intermittently attaching to the 
cold reservoir temperature, $T_{C}$ in between. Again the first mode of operation, the number of spin-half particles in the exited state is constant and the change of energy level increase from $\Delta_{c}$ to $\Delta_{6}$ in a very fast manner, after the system left to exchange heat with the cold reservoir and reached in thermal equilibrium, when state (${P_{C}^{e}, \Delta_{C}}$) detach from the cold
reservoir. During this sudden change of energy levels the system can allows extract work, this amount of work done on 
the system  defined as
\begin{equation}
 W_{6\rightarrow C} = P^{C}_{e}[\Delta_{6}-\Delta_{c}].
\end{equation}
Likewise, we can obtained the amount of work done by the system in the adiabatic processes 
($C\rightarrow D$) can be expressed as
\begin{equation}
\begin{split}
 W_{7\rightarrow8}& = P_{7}^{e}[\Delta_{8}-\Delta_{7}],
\text{ } \text{ } \text{ } \text{ and}\\
 W_{9\rightarrow10}& = P_{9}^{e}[\Delta_{10}-\Delta_{9}].
\end{split}
\end{equation}
The work done on the system  in the isothermal processes $C\rightarrow D$ is expressed by
\begin{equation}\label{WCD}
 W_{C\rightarrow D} = \Delta^{'}[P_{7}^{e} + P_{9}^{e} +P_{c}^{e}],
\end{equation}
where $\Delta{'} = \frac{\Delta_{D}-\Delta_{C}}{n}$, n is the number of subdivision.\\
\noindent In the second mode of operation, the change of energy level  $\Delta_{6}$ constant and the systems is allows to exchange heat until it relaxes back to thermal equilibrium at state $7$, when the system attach to the cold reservoir. 
During this process($6\rightarrow 7$), the systems is allowed to exhange spin-half particles in the exited state 
due to released the amount of heat, $Q_{6\rightarrow7}$
\begin{equation}
 Q_{6\rightarrow7}= \Delta_{6}[P_{7}^{e}-P_{6}^{e}],
\end{equation}
when it attaches to the cold reservoir. This lost heat decreases the number of spin-half particles from  the exited state 6($P^{6}_{e}$) to the exited state 7 ($P^{7}_{e}$). Correspondingly, 
we can obtained the amount of releavolumesed heat during in the 
isochoric processes of $C\rightarrow D$, which can be expressed as
\begin{equation}
\begin{split}
 Q_{8\rightarrow9}= \Delta_{8}[P^{9}_{e}-P^{8}_{e}],
\text{ } \text{ } \text{ } \text{ and} \text{ } \text{ } \text{ } 
 Q_{10\rightarrow D}= \Delta_{D}[P^{D}_{e}-P^{10}_{e}].
\end{split}
\end{equation}
The total heat released in the finite-time process 
of $C\rightarrow D$ becomes
\begin{equation}
 Q_{C\rightarrow D}= \Delta_{D}P_{D}^{e}-\Delta_{C}P_{C}^{e} - \Delta^{'}[P_{C}^{e}+P_{7}^{e}+P_{9}^{e}].
\end{equation}
In this finite-time process ($ C \rightarrow D $), the released heat is negative, where as work done on the system
is positive.
\subsection{Adiabatic branch of the process}
When the system reaches state B, (${P_{B}^{e}, \Delta_{B}}$), then it detached from the hot reservoir at a inverse temperature, $\beta_{1}$ and instantaneously connected to the cold reservoir at a  inverse temperature, $\beta_{2}$. Simultaneously, the energy level spacing of the system instantly changes from $\Delta_{B}$ to $\Delta_{C}$. However, the number of none-interacting spin-half particles in the excited state of the system remains unchanged (i.e $P_{B}^{e} =P_{C}^{e}$). During this branch of operation, there is no heat exchange between the system and the reservoirs. So that, the amount of work 
done by the system is described by
\begin{equation}\label{WCB}
W_{B\rightarrow C}= P_{B}^{e}[\Delta_{C}-\Delta_{B}]. 
\end{equation}
The other adiabatic process of the system occurs, when the system reaches state D (${P_{D}^{e}, \Delta_{D}}$),
then it detached from the cold reservoir at a temperature, $\beta_{2}$ and instantaneously connected to the hot reservoir at a inverse temperature, $\beta_{1}$. 
Simultaneously, the energy level spacing of the system instantly changes from $\Delta_{D}$ to $\Delta_{A}$
without change of the number of none-interacting spin-half particles in the excited state of the 
system (i.e $P_{D}^{e} =P^{e}_{A}$). The amount of work done on the system 
during this branch described by
\begin{equation}\label{WDA}
W_{D\rightarrow A}= P^{e}_{A}[\Delta_{A}-\Delta_{D}]. 
\end{equation}
After the end of adiabatic processes D to A  instantly the system returns back to initial state, ({$P^{D}_{e},\Delta_{A}$ }) in one completed cycle. \\

\noindent The net amount of heat and work done per cycle, respectvely given by 
\begin{multline}
Q_{net} = \Delta_{B}P_{4}^{e}-\Delta_{A}P^{e}_{A}+ \Delta [P^{e}_{A}+P_{2}^{e}+P_{4}^{e}]\\+\Delta_{D}P_{D}^{e}-\Delta_{C}P_{C}^{e} - \Delta^{'}[P_{C}^{e}+P_{7}^{e}+P_{9}^{e}].
\end{multline}
\begin{multline}\label{network}
 W_{net} = (\Delta_{C}-\Delta_{B})P_{B}^{e}[1+\frac{(\Delta_{A}-\Delta_{D})P^{e}_{A}}{(\Delta_{C}-\Delta_{B})P_{B}^{e}}] + \\
 \Delta^{'}P_{C}^{e} -\Delta P^{e}_{A} + (\Delta^{'}-\Delta)(P_{2}^{e} + P_{4}^{e}).
\end{multline}
\noindent The corresponding efficiency of the two-level quantum heat engine can be expressed as  the ratio of the net work done per cycle to 
the input heat per cycle, which is given by 
\begin{multline}\label{efficency1}
 \eta = \frac{(\Delta_{B}-\Delta_{C})P_{B}^{e}[1-\frac{(\Delta_{A}-\Delta_{D})P^{e}_{A}}{(\Delta_{B}-\Delta_{C})P_{B}^{e}}] } 
 {\Delta_{B} P_{B}^{e}-\Delta_{A} P^{e}_{A}+\Delta(P^{e}_{A}+P_{2}^{e}+P_{4}^{e})}.\\
 -\frac{Delta^{'}P_{C}^{e} +\Delta P^{e}_{A} - (\Delta^{'}-\Delta)(P_{2}^{e} + P_{4}^{e})}
 {\Delta_{B} P_{B}^{e}-\Delta_{A} P^{e}_{A}+\Delta(P^{e}_{A}+P_{2}^{e}+P_{4}^{e})}.
\end{multline}
The efficiency (described in Eq. \eqref{network}) of a two-level quantum engine cycle process depends on
probabilities of the ground and exited state of the system and the change of energy levels.\\
\noindent The Power of the two-level quantum heat engine can be expressed as,
\begin{equation}\label{powr}
 P(t) = -\frac{W_{net}}{t}.
\end{equation}
The negative sign in Eq.\eqref{powr} expressed that work is done by the engine, because power is extract or something needs get out of the system. Then, inserting Eqs. \eqref{network} into \eqref{powr}, the power of the two-level quantum heat engine can be rewritten as,
\begin{multline}
\label{powers}
 P(t) = \frac{ (\Delta_{B}-\Delta_{C})P_{B}^{e}[1-\frac{(\Delta_{A}
 -\Delta_{D})P^{e}_{A}}{(\Delta_{B}-\Delta_{C})P_{B}^{e}}]}{t}\\-\frac{\{(\Delta_{D}-\Delta_{C})P_{C}^{e} -(\Delta_{A}-\Delta_{B}) P^{e}_{A}\}}{t^2}\\ 
 -\frac{\{(\Delta_{D}-\Delta_{C}-\Delta_{A}+\Delta_{B} (P_{2}^{e} + P_{4}^{e})\}}{t^{2}},
\end{multline}
where $\Delta = \frac{\Delta_{A}-\Delta_{B}}{n}$ and $\Delta{'} = \frac{\Delta_{D}-\Delta_{C}}{n}$, n is the number of subdivision i.e.$ n\propto t$. 
\noindent Furthermore, Eqs. \eqref{efficency1} and \eqref{powers} show that the efficiency and power output are both as a function of $P^{n}_{b}$ and $\Delta_{b}$ . Therefore, we can generate the curves of power output with respect to the efficiency by varying $P^{n}_{b}$ and $\Delta_{b}$ .
\begin{figure}[h]
 \centering
 \includegraphics[width=7cm,height=7cm]{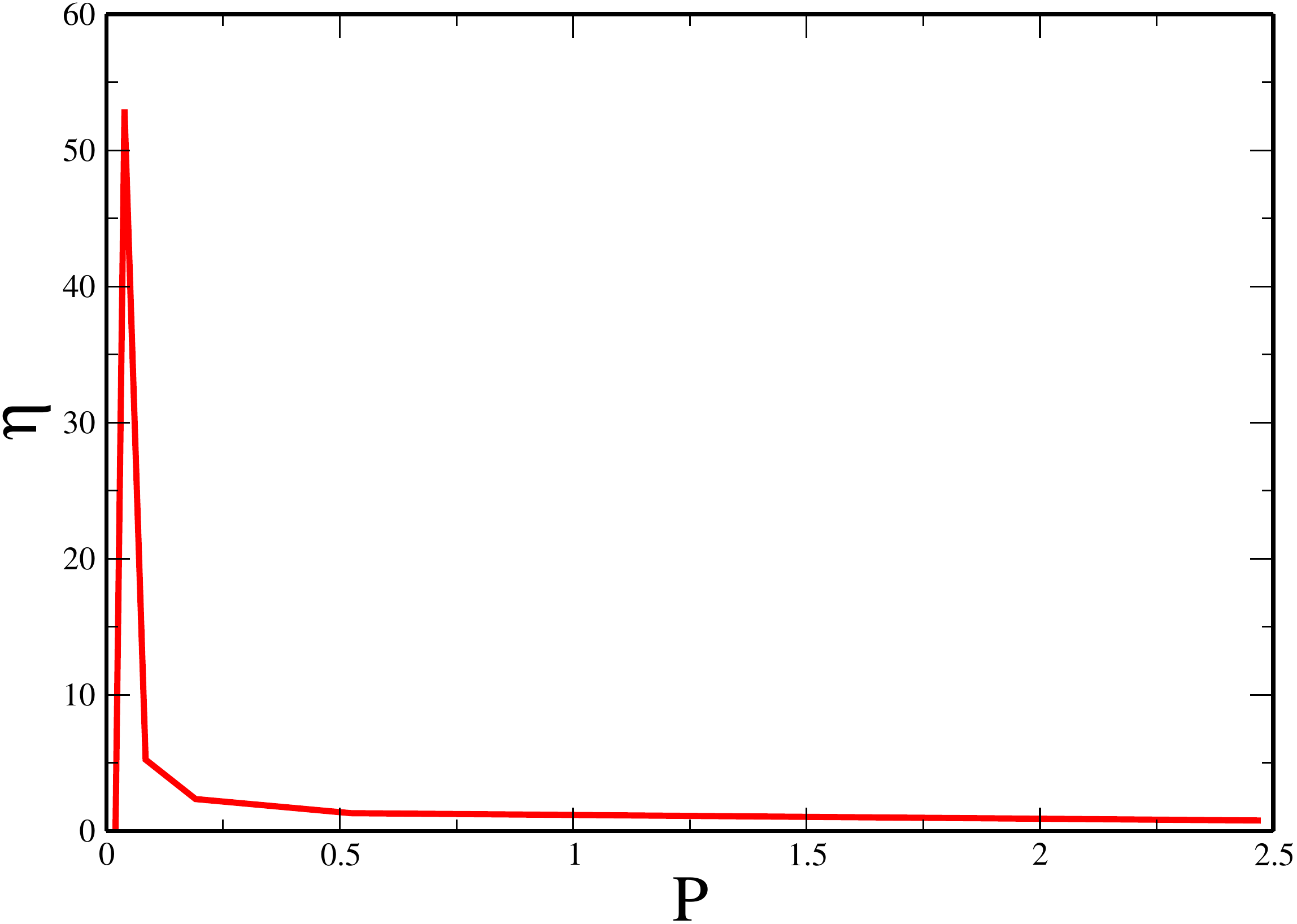}
 \caption{The plot of efficiency  vs  power of the model, when $\beta_{1} = 0.8, \beta_{2} =1.5,  P_a= 0.2,  P_b= 0.34, 
P_2=0.28$ and $ P_4 =0.32$}
 \label{kkk}
\end{figure}
\noindent In Fig. \eqref{kkk} shows the plot of efficiency as a function of power output. The efficiency increases with the increasing of power output. At an optimal value of power, the efficiency becomes maximum and monotonically decreasing with increasing power. It means that in order to improve the engine’s efficiency, the cost is to decrease the engine’s power output, and vice versa. Therefore, this kind of trade-off between the efficiency and power output should be concerned when the engine is working at the lower bound of the region.
\section{Efficiency at maximum power of the model}\label{section5}
\noindent In this section, we study the efficiency at maximum power of the model and compare with that of the Curzon-Alborn efficiency. In order to maximize the output power of the model with respect to the time, we need to solve
\begin{equation}\label{maxp}
 {\frac{\partial{P(t)}}{×\partial{t}}}\mid_{{t}_{mp}}=0.
\end{equation}
Then, using Eqs. \eqref{maxp} and \eqref{powers}, we get
\begin{equation}
 t_{mp} = \frac{2\Gamma}{×\chi},
\end{equation}
where $ \chi= (\Delta_{B}-\Delta_{C})P^{B}_{e}[1-\frac{(\Delta_{A}-\Delta_{D})P^{e}_{A}}{(\Delta_{B}-\Delta_{C})P^{B}_{e}}]$ and
$\Gamma= (\Delta_{D}-\Delta_{C})P^{C}_{e} -(\Delta_{A}-\Delta_{B}) P^{e}_{A} + (\Delta_{D}-\Delta_{C}-\Delta_{A}+\Delta_{B})(P^{2}_{e} + P^{4}_{e})$. 
The efficiency at maximum power ($\eta_{mp}$) of the two-level quantum heat engine becomes
\begin{equation}\label{mpeef}
 \eta_{mp} = \frac{\eta_{C}}{2} \frac{1}{1+ \frac{\eta_{C}}{2} \frac{\varphi }{{\Gamma}}} ,
\end{equation}
where 
 $\varphi= (\Delta_{A}-\Delta_{B})(P^{e}_{A}+P^{2}_{e}+P^{4}_{e})$.

\begin{figure}[h]
 \centering
 \includegraphics[width=7cm,height=7cm]{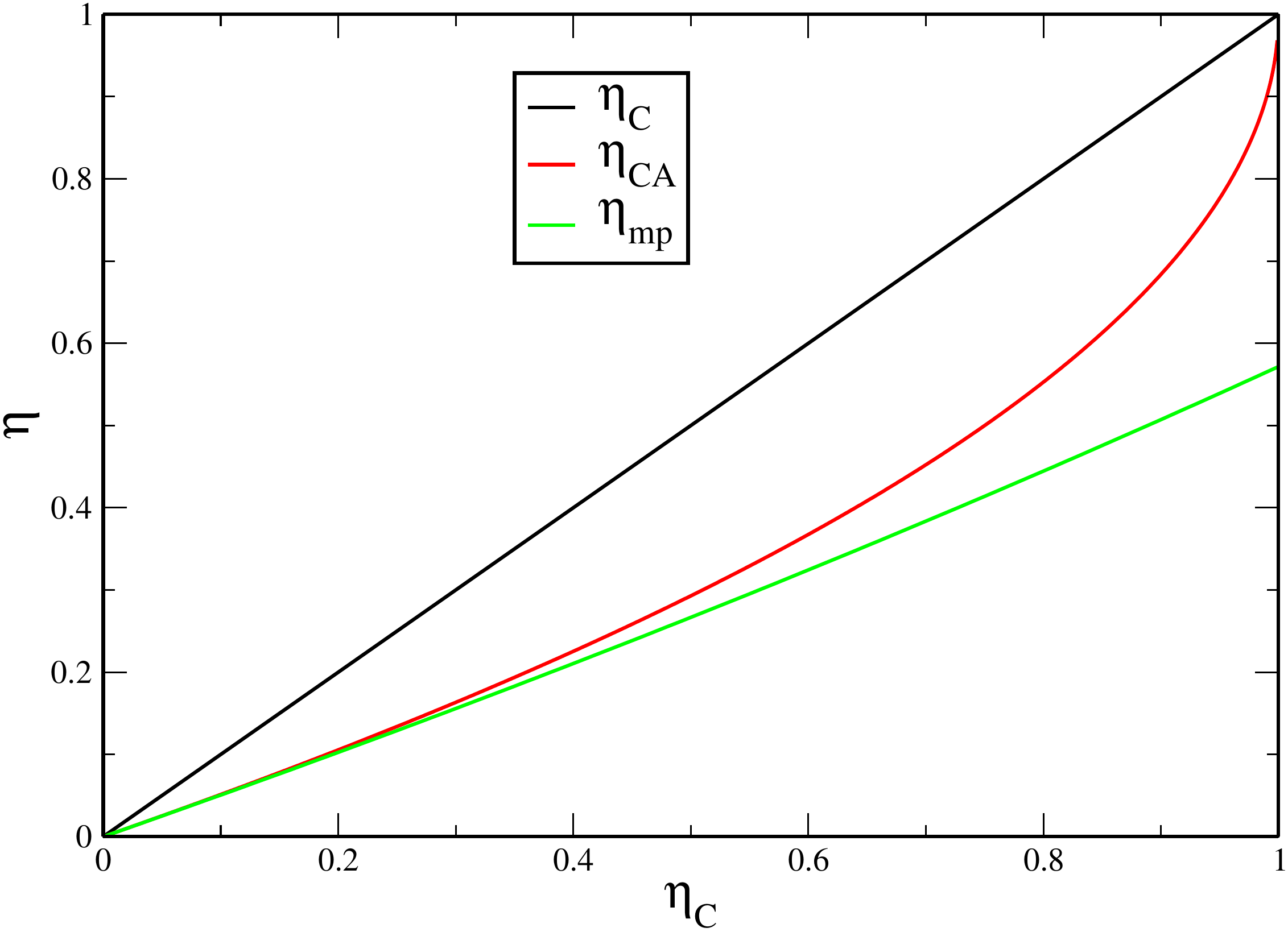}
 \caption{The plot of efficiency at maximum power of the model vs Carnot efficiency}
 \label{maximumpower}
\end{figure}
\noindent In Fig. \eqref{maximumpower}, we plot the efficiency at maximum power as a functions of $\eta_{C}$ with the parameter choice $\Gamma=-0.070$ and $\varphi=0.088$. In the linear regime, our result agree with the $\eta_{CA}$. But in the higher order of $\eta_{C}$, the efficiency at maximum power lies below the $\eta_{CA}$.\\
The efficiency at maximum power of Eq.\eqref{mpeef} of the model can be rewritten as  
\begin{equation}
 \eta_{mp}= \frac{ \eta_{c}}{2} - \frac{\varphi\eta^{2}_{c}}{4\Gamma}+ \frac{\varphi^{2}\eta^{3}_{c}}{8\Gamma^{2}}
 -\frac{\varphi^{3}\eta^{4}_{c}}{16\Gamma^{3}}+\vartheta(\eta^{5}_{c}).
\end{equation}
This expression result agree with the universal value for the efﬁciency at maximum power \cite{schmiedl2008efficiency,tu2008efficiency, esposito2009thermoelectric, allahverdyan2008work, izumida2012efficiency}
similar to that of Carnot efﬁciency of the heat engines, when $\frac{\varphi}{\Gamma} = -0.5$ value.
\section{Summary and Conclusion}\label{section6}
\noindent In this work, we have taken a simple model of a two-level quantum heat engine. The spin-half particle consider as a working substance in the presence of external magnetic field. We have investigated the thermodynamic properties of the two-level quantum heat engine in the fine time condition such as heat, power and efficiency per cycle. The work per engine cycle takes the maximum value in the limit where the time intervals of the isothermal processes tend to infinity. The efficiency at maximum power evaluated and the coefficient of the
linear term for the efficiency at maximum power is $\frac{1}{2}$.\\
\noindent In conclusion, the result of our model agrees with the universality value of the efficiency at maximum
power up to the first order in $\eta_{c}$ while in the higher order of $\eta_{C}$, the efficiency at maximum power lies below the $\eta_{CA}$.\\
\noindent \textit{Acknowledgments:  Y.B. would like to thank  Wolkite University for financial support during his work.} 
\section*{Author contributions}
YB conception and design of study. YB and TB performed the analytic  calculations and numerical results. YB, TB,YA, and  AA analysing and interpreting the results. YB, TB, YA, and AA drafted manuscript preparation. All authors reviews the results and approved the final version of the manuscript. \\
\textbf{Funding}: The authors declare that they have no known competing financial interests.\\
\textbf{Conflict of Interest}: The authors declare no conflict of interest.\\

\section*{Data Availability Statement}
This manuscript has no associated data or the data will not be deposited.

\bibliographystyle{spphys}
\bibliography{mybibfile.bib}

\end{document}